\documentclass[draft]{agujournal2019}
\usepackage{url} 
\usepackage{lineno}
\usepackage[inline]{trackchanges} 
\usepackage{soul}
\usepackage{comment}

\draftfalse

\journalname{JGR: Space Physics}

\begin{document}

\title{Extent of the magnetotail of Venus from the Solar Orbiter, Parker Solar Probe and BepiColombo flybys}

\authors{Niklas J.T. Edberg\affil{1}, David J. Andrews\affil{1}, J. Jordi Bold{ú}\affil{1,2}, Andrew P. Dimmock\affil{1}, Yuri V. Khotyaintsev\affil{1,2}, Konstantin Kim\affil{1,2}, Moa Persson\affil{1}, 
Uli Auster\affil{3}, Dragos Constantinescu\affil{3}, Daniel Heyner\affil{3} , Johannes Mieth\affil{3}, Ingo Richter\affil{3},
Shannon M. Curry\affil{4},  
Lina Z. Hadid\affil{5}, 
David Pisa\affil{6}, 
Luca Sorriso-Valvo\affil{1,7,8}, 
Mark Lester\affil{9}, Beatriz S\'{a}nchez-Cano\affil{9}, Katerina Stergiopoulou\affil{9},
Norberto Romanelli\affil{10,11},
David Fischer\affil{12}, Daniel Schmid\affil{12}, Martin Volwerk\affil{12}}

\affiliation{1}{Swedish Institute of Space Physics, Uppsala, Sweden}
\affiliation{2}{Department of Physics and Astronomy, Uppsala University, Sweden}
\affiliation{3}{Institut für Geophysik und extraterrestrische Physik, Technische Universität Braunschweig, Braunschweig, Germany}
\affiliation{4}{Laboratory for Atmospheric and Space Plasmas, University of Colorado, Boulder, CO, USA}

\affiliation{5}{LPP, CNRS, Observatoire de Paris, PSL Research University, Sorbonne Université, École Polytechnique, Institut Polytechnique de Paris, 91120 Palaiseau, France}

\affiliation{6}{Dept. of Space Physics, Institute of Atmospheric Physics of the Czech Academy of Sciences, Prague, Czechia}

\affiliation{7}{CNR/ISTP – Istituto per la Scienza e Tecnologia dei Plasmi, Via Amendola 122/D, 70126 Bari, Italy}
\affiliation{8}{Division of Space and Plasma Physics, KTH Royal Institute of Technology, Stockholm, Sweden}

\affiliation{9}{School of Physics and Astronomy, University of Leicester, Leicester, UK}
\affiliation{10}{NASA Goddard Space Flight Center, Greenbelt, MD, USA}

\affiliation{11}{University of Maryland College Park, College Park, MD, USA}
\affiliation{12}{Space Research Institute, Austrian Academy of Sciences, Graz, Austria}
\correspondingauthor{Niklas J. T. Edberg}{ne@irfu.se}


\begin{keypoints}
\item Venus' magnetotail is observed \textcolor{black}{during} nine spacecraft flybys revealing a dynamic structure reaching at least 60 R$_V$ downstream.
\item An improved bow shock model is presented for the deep tail region. 
\item The pre-existing model of the induced magnetospheric boundary is valid downstream to at least 20 R$_V$.
\end{keypoints}

\begin{abstract}
We analyze data from multiple flybys by the Solar Orbiter, BepiColombo, and Parker Solar Probe missions to \textcolor{black}{study the interaction between Venus' plasma environment and the solar wind forming the induced magnetosphere. Through examination of magnetic field and plasma density signatures we characterize the spatial extent and dynamics of Venus' magnetotail, focusing mainly on boundary crossings.} Notably, we observe significant differences in boundary crossing location and appearance between flybys, highlighting the dynamic nature of Venus' magnetotail.  \textcolor{black}{In particular, during Solar Orbiter’s third flyby, extreme solar wind conditions led to significant variations in the magnetosheath plasma density and magnetic field properties, but the increased dynamic pressure did not compress the magnetotail. Instead, it is possible that the increased EUV flux at this time rather caused it to expand in size. Key findings also include the identification of several far downstream bow shock, or bow wave, crossings to at least 60 R$_V$ (1 R$_V$ = 6052 km is the radius of Venus), and the induced magnetospheric boundary to at least $\sim$20 R$_V$}. These crossings provide insight into the extent of the induced magnetosphere. Pre-existing models from Venus Express were only constrained to within $\sim$5 R$_V$ of the planet, and we provide modifications to better fit the far-downstream crossings. \textcolor{black}{The new model BS is now significantly closer to the central tail than previously suggested, by about 10 R$_V$ at 60 R$_V$ downstream.}
\end{abstract}

\section{Introduction}\label{section:introduction}
Venus and its atmosphere, ionosphere, and induced magnetosphere have been studied with several space missions in the past decades. The investigations will continue in the coming years with new dedicated Venus missions, although most will not carry any plasma instruments. While the near-Venus plasma environment has been studied extensively by orbiting spacecraft such as Pioneer Venus Orbiter (PVO), Venus Express (VEX) and Akatsuki, the far tail is less well known. 

In recent years, starting in 2018, Venus has been subject to numerous flybys by the Solar Orbiter, BepiColombo, and Parker Solar Probe (PSP) spacecraft \cite{muller2020, benkhoff2021, fox2016}. Solar Orbiter and PSP are both missions designed to study the solar wind and the Sun while BepiColombo is aimed at Mercury, and all three use Venus for gravity assist manoeuvres to steer the spacecraft into their designated trajectories. Solar Orbiter has to date completed three flybys, BepiColombo two, and PSP six with details of these flybys listed in Table \ref{times}. These measurements were not strictly unique for Venus, since some passes through the tail were also made by Mariner 5 \& 10 and Venera 4,6,9 \& 10 in the 1960's and 1970's, see e.g. \citeA{slavin1984}, and Galileo in 1990 \cite{kivelson1991}. Cassini also flew past Venus twice, in 1998 and 1999 \cite<e.g.>{gurnett2001}, but did not provide sufficient sampling of the far tail. \textcolor{black}{PVO was orbiting Venus, but it should be mentioned that its apoapsis reached around 11 R$_V$ downstream and so could perform measurements in the tail. Finally, Messenger also made a flyby of Venus in 2007 on its way to Mercury and crossed the bow shock outbound at approximately 5 R$_V$ downstream \cite{slavin2009}.}
\begin{table*}
    \centering
    \caption{Summary of the Venus flybys used in this paper, their closest approach (C/A) altitude and times of C/A. Note that Solar Orbiter will complete another five flybys while PSP has already completed six out of seven flybys, but none of these additional flybys will pass through the far tail of Venus and so are not included in our study.}
    \begin{tabular}{c|c|c}
        Flyby & C/A [R$_V$] & Time of C/A  \\
        \hline
        Solar Orbiter 1 & 1.23 & 2020-12-27T12:39 \\
        Solar Orbiter 2 & 1.32 & 2021-08-09T04:41 \\
        Solar Orbiter 3 & 1.06 & 2022-09-04T01:26 \\
        BepiColombo 1   & 1.77 & 2020-10-15T03:58 \\
        BepiColombo 2   & 0.09 & 2021-08-10T13:52 \\
        PSP 1           & 0.40  & 2018-10-03T08:44 \\
        PSP 2           & 0.50  & 2019-12-26T18:15 \\
        PSP 3           & 0.14 & 2020-07-11T03:24\\
        PSP 4           & 0.39 & 2021-10-16T20:06 \\
        \hline
    \end{tabular}    
    \label{times}
\end{table*}

The geometry of many of these recent flybys is such that they pass through the bow shock (BS) and magnetotail far downstream ($\sim10-100$ R$_V$, where 1 R$_V$ = 6052 km is the radius of Venus), \textcolor{black}{and thereby enable studies of the properties of the tail, and in particular the plasma boundaries at these distances}. From several of the individual flybys, studies have already been undertaken to characterize the magnetotail  electric and magnetic fields, plasma waves, and energetic particles residing there \cite{hadid2021, allen2021, volwerk2021, volwerk2021b}. \citeA{dimmock2022} investigated the detailed structure and dynamics of the dayside BS using the Solar Orbiter high-time resolution measurements of electric and magnetic fields, while \citeA{aizawa2022} and \citeA{stergiopoulou2023} performed model-data comparisons from the second BepiColombo and the first two Solar Orbiter flybys, respectively, to gain a deeper understanding of the solar wind interaction with Venus. \citeA{bowen2021} used high-cadence electric field measurements from PSP to show the existence of kinetic-scale turbulence present in the tail of Venus. Furthermore, \citeA{persson2022} used BepiColombo data to determine the extent of the previously unsampled stagnation region upstream of Venus, while \citeA{collinson2021, collinson2022} investigated the ionospheric density structure and the near-Venus magnetotail structure from PSP. However, so far there has been no attempt to combine these flyby measurements to perform a statistical study of the plasma environment around Venus, and in particular of the far tail plasma boundaries arising from the interaction between the solar wind and Venus.

Nomenclature of plasma boundaries at unmagnetized planets, such as Venus and Mars, is a delicate matter. Various names have been suggested for similar discontinuities and boundaries when observed by different instruments, on different missions. In this paper, we will use the terms bow shock and induced magnetospheric boundary (IMB). Close to the planet, the supersonic solar wind is obstructed by the planet and its ionosphere leading to a deceleration to subsonic speed, whereby a shock is formed. Farther downtail, where the flow is more tangential to the induced magnetosphere and the magnetosheath plasma flow has increased to a large fraction of the solar wind speed, the deceleration is not necessarily abrupt enough to form a proper shock. For simplicity, we will use the term BS far downstream too even though bow wave might be a more appropriate name. 

In this study, we will use a combination of the following criteria for identifying the BS when moving inbound: an abrupt increase in magnetic field strength and density; an increase of magnetic field fluctuations; a rotation of magnetic field (used farther downtail), and for the IMB: a gradual increase (pile-up) of magnetic flux; a transition to a region of lower density; a decrease of magnetic fluctuations, a rotation of the magnetic field direction. Not all criteria need to be met for all boundary crossings, and especially not farther downtail where the transition from one region to another need not be abrupt. A change in electron and ion distributions and changing ion compositions across the boundaries could also have been used for identification, if such measurements would have been available. More thorough descriptions of the plasma boundaries and different naming conventions can be found in \citeA{phillips1991, bertucci2011, futaana2017}.

We will use the Venus-centered Venus Solar Orbital (VSO) frame to describe the orbits and structure of the plasma environment. In the VSO frame, the $X$-axis points from Venus to the Sun, the $Z$-axis is directed normal to the orbital \textcolor{black}{plane pointing northward, and the $Y$-axis is approximately anti-}parallel to the orbital velocity vector of Venus and completes the right-handed system. BSs around planets and comets in the solar system, and the boundaries beneath, and in particular their location and shape have been studied extensively over the years \cite{slavin1984, huddleston1998, masters2008, edberg2008, martinecz2009, edberg2023}. \citeA{martinecz2008, martinecz2009} used several years of Venus Express data to compile an empirical model of both the BS and the IMB (called upper mantle boundary in those studies) and compared it to previous studies from Pioneer Venus Orbiter. Most empirical and numerical models use conic sections to fit the BS, of the form

\begin{equation}
r=\frac{L}{1+\epsilon\ \cos \theta},
\label{conicsection}
\end{equation}
where $L$ is the semi-latus rectum and $\epsilon$ is the eccentricity. ($r$, $\theta$) are the polar coordinates in an aberrated frame VSO$'$ $(X', Y', Z')$ \textcolor{black}{centered at ($X_0, 0, 0$). To account for the aberration, i.e. that the planet's transverse velocity $v_{y}=35$ km/s is non-negligible compared to the solar wind radial velocity $v_{r}=400$ km/s, the VSO$'$ frame is created by rotating the VSO frame counter-clockwise about the $Z_{VSO}$-axis with an angle $arctan(v_y/v_r) = 5^\circ$, such that the solar wind's apparent direction is in the $-X'$-direction. This correction} is especially important for crossings far downtail as a small rotation gives a large lateral change. 

A conic section model is what \citeA{martinecz2008, martinecz2009} used for the BS, while the IMB shape was represented by a circle on the dayside and a straight line on the nightside \cite{martinecz2009}. More recently, \citeA{signoles2023} revisited the topic of shape and location of the BS and used a larger set of Venus Express data to update the model parameters of \citeA{martinecz2009} for the BS. \citeA{whittaker2010} also used a conic section to model the dayside BS, but a straight line on the night side, representing a Mach cone at an angle of 10.5$^\circ$ to the VSO$'$ X$'$-axis. Mach cones angles at various planets have also been investigated by \citeA{slavin1984}, who gave an observational value of the angle of 13.9$^\circ$, for a magnetosonic Mach number of 6.6. \citeA{zhang2008d} also found that an angle of 10.5$^\circ$ fitted their smaller number of identified BS crossings well. It is important to note that the large statistical BS and IMB models of \citeA{martinecz2009, whittaker2010, signoles2023} were all confined to within 5 R$_V$, i.e., the orbital limit of Venus Express. PVO reached about twice that distance while Mariner 5 \& 10 and Venera 4,6 9 \& 10 provided an additional handful of crossings down to 15 R$_V$, which \citeA{slavin1984} used to derive a BS model. While \citeA{signoles2023} used the hitherto largest data set of BS crossings, \citeA{slavin1984} used crossings farther downstream with data from the early Mariner and Venera missions, to constrain their BS model to 15 R$_V$ downstream.
 
\citeA{whittaker2010} and \citeA{shan2015} found a clear variation in BS stand-off distance with extreme ultraviolet (EUV) solar emission variations, confirming previous results from the Pioneer Venus Orbiter (PVO) era \textcolor{black}{\cite{russell1988, alexander1985}}. The BS at the subsolar point moved from 1.364 R$_V$ during solar minimum to 1.459 R$_V$ at solar maximum, while the terminator shock distance moved from 2.087 R$_V$ to 2.146 R$_V$, i.e., about a 7\% and 3\% change, respectively \cite{shan2015}. \citeA{signoles2023} also studied the factors influencing the BS and IMB (called ion composition boundary in that paper). They found that the BS is mainly dependent on the EUV flux, but also on the interplanetary magnetic field (IMF) magnitude, the Alfvén Mach number, and the solar wind dynamic pressure during solar maximum. The IMB was found to mainly move with changes in the dynamic pressure \textcolor{black}{and IMF magnitude (i.e. magnetic pressure). These trends were also more pronounced during solar maximum conditions}. 

In this paper, we will provide further information on the structure of the far magnetotail by using the measurements from the flybys by Solar Orbiter, BepiColombo, and PSP. We will later compare our results primarily with the model of \citeA{signoles2023}, which was not very dissimilar to the model of \citeA{slavin1984} in the far tail (at 60R$_V$ the two models differ by 1.5 R$_V$ in width). 

The paper is structured as follows: in Section \ref{section:data}, we describe the flyby trajectories as well as the instruments and data used. In Section \ref{section:observations} we describe the observations from each mission during their respective flyby, and in \ref{section:model} we provide a new BS model as well as test the validity of the IMB model. This is followed by a discussion in Section \ref{section:discussion} and our conclusions in Section \ref{section:conclusion}. 

\begin{figure*}
\includegraphics[width=16cm]{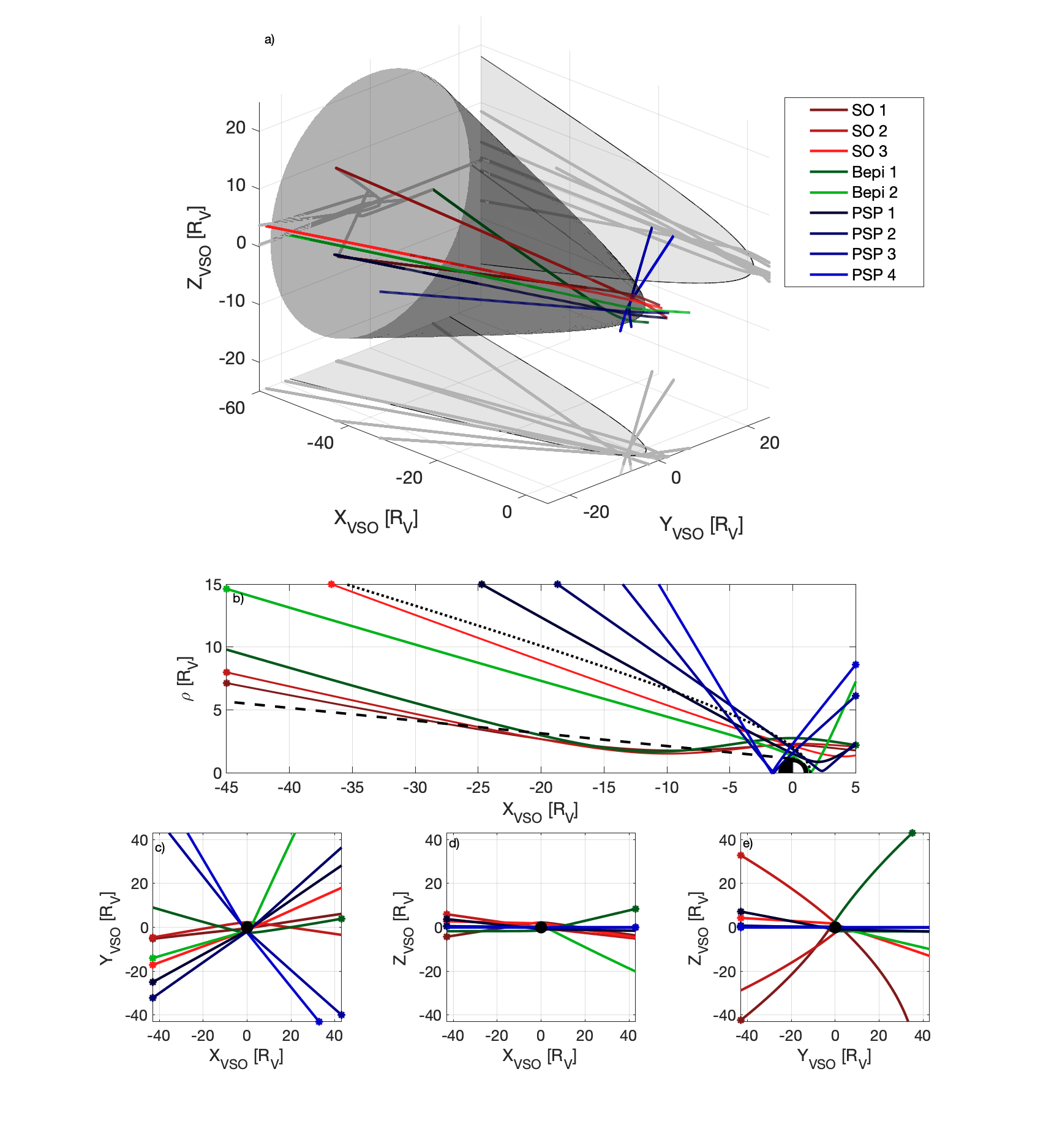}
\caption{(a) Trajectory overview of the Solar Orbiter (red), BepiColombo (green), and PSP (blue) flybys of Venus, in VSO coordinates. A rotationally symmetric BS model based on the work of \citeA{signoles2023} is included (transparent grey color) \textcolor{black}{together with projections of the trajectories and the BS (light grey lines). (b) The trajectories, the BS model from \citeA{signoles2023} (dotted line), and IMB model from \citeA{martinecz2009} (dashed line), shown in cylindrical VSO coordinates. The stars at the edges indicate the inbound legs of each pass. (c-e) The projections of the trajectories on the three planes X-Y, X-Z, and Y-Z.}}
\label{3d}
\end{figure*}
\section{Instruments, Data and Trajectories}\label{section:data}
In this paper, we primarily use measurements from fluxgate magnetometers on three spacecraft: on Solar Orbiter (MAG) \cite{horbury2020}, on the Mercury Planetary Orbiter of BepiColombo (MPO-MAG)\cite{glassmeier2010, heyner2021}, and the magnetometer within the FIELDS suite on PSP \cite{bale2016}. Additionally, we utilize data from the Radio and Plasma Wave (RPW) instrument on Solar Orbiter \cite{maksimovic2020}, which provides a high cadence measure of the electron density through the measured probe-to-spacecraft potential cross-calibrated to the density obtained from the plasma frequency line \cite{khotyaintsev2021}.The magnetic field and density data are resampled to 1 Hz in this paper. 

Solar Orbiter carries particle instruments as part of its nominal mission but these were turned off during the Venus flyby for safety reasons, except for the higher energy particle instrument \textcolor{black}{\cite{allen2021,wimmer2021}.} BepiColombo also carries several particle sensors \textcolor{black}{and some measurements do exist from the flybys \cite{persson2022, aizawa2022, rojo2024}. Particularly the electron spectrometer (MEA) data would have been useful for studying bowshock crossing, but since not all energy steps were used and not more than one or two pixels had a field of view out of the MOSIF shield, there is little confidence in those measurements in this interval.}

Figure \ref{3d} shows the trajectories of the three spacecraft during their Venus flybys. Solar Orbiter approached Venus from the far tail during its three passes, while BepiColombo approached from upstream during the first flyby and from the tail during the second flyby. PSP has completed six flybys to date but only four of them have trajectories relevant for this study \textcolor{black}{-- on the 5th and 6th flyby PSP passed on the dayside so \textcolor{black}{these are} not used. PSP crossed the BS closer to the planet than Solar Orbiter and BepiColombo}, and approached from the tail during the first two flybys and from the upstream direction during the following two flybys. The first four PSP flybys entered the tail not much farther downstream than the long-lasting missions of PVO and VEX, but still provided some interesting complementary measurements of the tail structure. During PSP flyby 3 and 4 the Venus plasma environment was sampled downtail to about 7 and 5 R$_{V}$, respectively. 

\section{Observations}\label{section:observations}
As mentioned, many of these flybys traverse a region poorly studied in the past - the far tail of Venus' induced magnetosphere. Both Solar Orbiter and BepiColombo had trajectories that potentially could pass through many 10's to 100's of R$_{V}$ of Venus's tail, corresponding to many hours of measurements. \textcolor{black}{Interestingly, \citeA{grunwaldt1997} made observations of an ion beam originating from Venus 7400 R$_V$ downstream of the planet. How far a BS or any other permanent boundary or signature arising} from the solar wind interaction with Venus actually extends downstream has so far been unknown.

In the following section, we will present the data obtained during the flybys by Solar Orbiter, BepiColombo, and PSP, and in particular identify the crossings of the BS and IMB. We summarize the boundary crossing times in Table \ref{crosstab}.

\subsection{Solar Orbiter}
\begin{figure*}
\includegraphics[width=15cm]{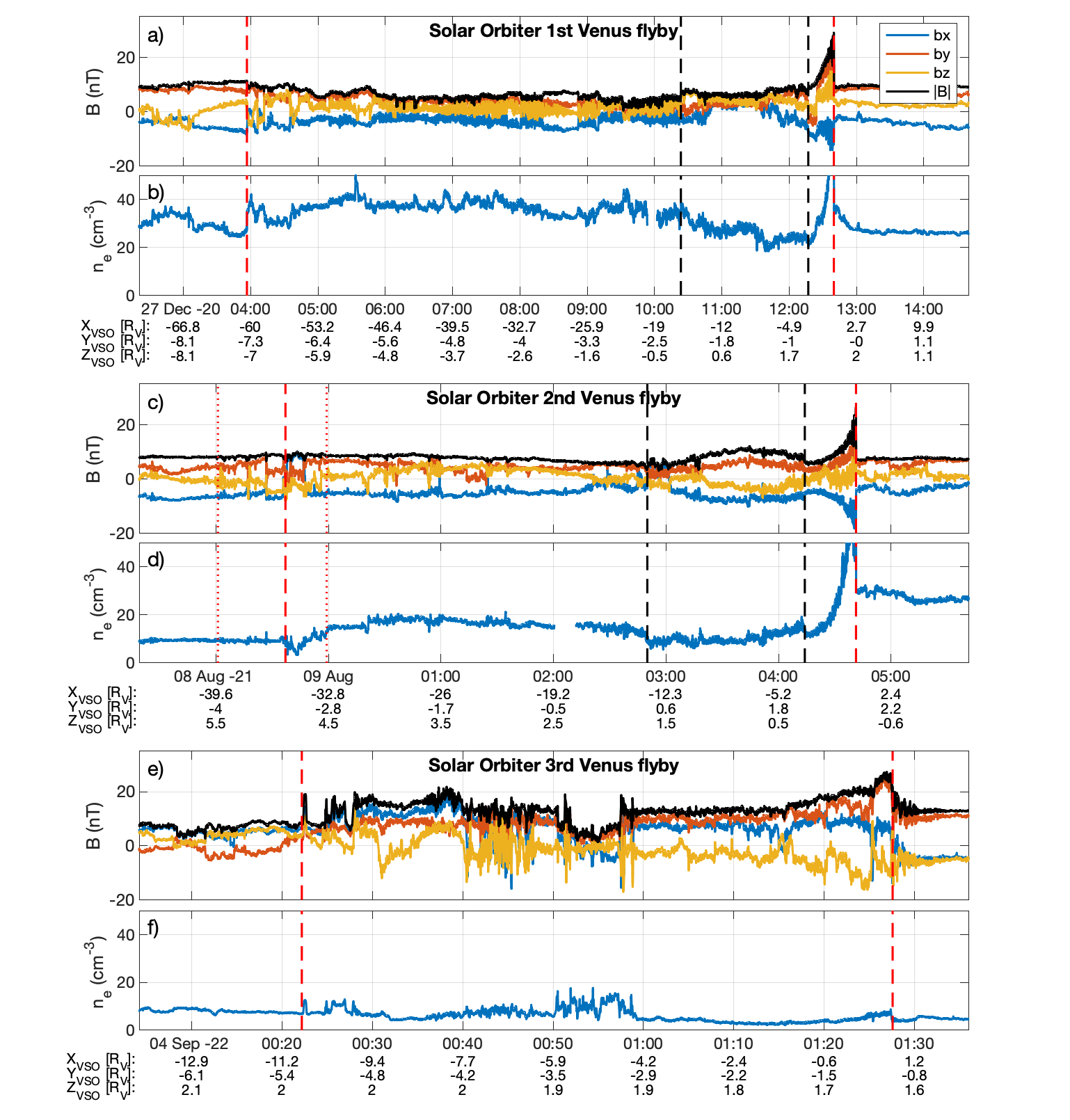}
\caption{Time series of Solar Orbiter magnetic field and electron density data from the first three Venus flybys. The entries and exits to and from the Venus plasma environment (BS or bow wave crossings) are indicated by the red dashed lines, and the crossings of the IMB by black dashed lines. On the ramside, a clear BS is developed, while farther downtail the transition becomes less clear, which is for the second flyby (panel c and d) indicated by the dotted red lines spanning an uncertainty interval.} 
\label{3venus}
\end{figure*}

Figure \ref{3venus} shows a time series of Solar Orbiter measurements of the magnetic field and electron density during its three flybys. Solar Orbiter entered Venus' induced magnetosphere from the tail and exited by crossing the BS close to the planet in all three cases. The entry, or transition, into the plasma environment far downtail is not obvious or abrupt in all cases, as the variations seen in both density and magnetic field at this distance are similar to those normally seen in the pristine and varying solar wind. However, a comparison between the three flybys and the presence of common features provide some leverage to the identification of boundaries.

During Solar Orbiter's first flyby (Figure \ref{3venus}a,b), a sudden magnetic field reversal can be seen at 03:56 UTC on 27 Dec 2020, accompanied by a sharp increase in plasma density, which is likely the transition into the Venus plasma environment \cite{stergiopoulou2023}. Following this, a long stretch of higher magnetic field fluctuations and increased plasma density lasts until the closest approach at 12:39 UTC and the clear outbound BS crossing at 12:40 UTC. The different plasma waves observed in this interval were presented by \citeA{hadid2021}. Before the outbound BS crossing, the tailside IMB is also traversed inbound and outbound. At 10:23 UTC, the field strength increases, a rotation of the field occurs, and the density starts to decrease. At 12:17 UTC, the spacecraft crosses the IMB outbound and moves into the dayside magnetosheath.
\begin{table*}
    \centering
     \caption{Summary of the times of all BS and IMB crossings by Solar Orbiter, BepiColombo \textcolor{black}{and PSP. Dates within parenthesis indicate uncertain crossings, as discussed in the text.}}
    \begin{tabular}{c|c|c|c|c|c}
                        & BS in  & IMB in  &  IMB out & BS out &\\
                         \hline
        SO 1 & 2020-12-27T03:56:40 & 2020-12-27T10:23:30 & 2020-12-27T12:17:00 & 2020-12-27T12:39:50 & \\
        SO 2 & (2021-08-08T23:37:00) & 2021-08-09T02:50:00 & 2021-08-09T04:14:00 & 2021-08-09T04:41:20 & \\
        SO 3 & 2022-09-04T00:22:10 & - & - & 2022-09-04T01:27:35 & \\
        Bepi 1 & 2020-10-15T04:07:30 & 2020-10-15T07:40:00 & (2020-10-15T13:00:00) & (2020-10-16T01:43:00) & \\
        Bepi 2 & (2021-08-10T09:43:00) & 2021-08-10T12:06:00 & 2021-08-10T13:06:00 & 2021-08-10T14:00:00 & \\
        PSP 1  & 2018-10-03T08:22:25 & - & - & - & \\
        PSP 2  & 2019-12-26T18:08:20 & - & - & 2019-12-26T18:13:40 & \\
        PSP 3  & 2020-07-11T03:18:20 & 2020-07-11T03:21:00 & 2020-07-11T03:35:00 & 2020-07-11T03:59:20 & \\
        PSP 4  & 2021-02-20T19:58:30 & 2021-02-20T20:04:00 & 2021-02-20T20:15:00 & 2021-02-20T20:34:20 & \\
        \hline
    \end{tabular}
    \label{crosstab}
\end{table*}

Solar Orbiter's second flyby (Figure \ref{3venus}c,d) had a trajectory very similar to the first, but the entry into the Venus plasma environment is less clear, at least in the magnetic field data. At 23:37 UTC on 21 Aug 2021, there is a sudden rotation in the field, but the field fluctuations following this are not as pronounced as during the first flyby. However, the density measurements show, similarly to the first flyby, a long stretch of increased values starting at this time. It is possible that there are signatures of an entry already half an hour earlier (around 23:\textcolor{black}{00} UTC) when the field starts to rotate and fluctuate. As \citeA{stergiopoulou2023} also indicated, it is not clear at what exact point the boundary is crossed if looking at the magnetic field data alone during this flyby. One could also argue that the magnetic field rotation at 23:37 UTC could simply be a solar wind structure such as a current sheet, but BepiColombo was at this time acting as a solar wind monitor about 1 hour downstream of Venus. Since they did not observe this structure passing by in the solar wind (see Figure \ref{comp} further below) it is likely that we do see a spatial change at this point, interpreted as a boundary crossing. \textcolor{black}{Nevertheless, we include here an uncertainty interval of about one hour (dotted red line in Figure \ref{3venus}b) to indicate that the boundary was not abrupt. This interval is chosen arbitrarily but meant to be rather generous around the possible boundary to make sure that the plasma regime changes across the interval.} Outbound, Solar Orbiter clearly crossed the BS at 04:41 UTC on 9 Aug 2021. Similar to the first flyby, the IMB was also traversed inbound and outbound, at 02:50 and 04:14 UTC, respectively, indicated by the increase in field strength, rotation of the field direction, and an overall decrease in density. However, the outbound IMB crossing shows an abrupt density decrease before increasing again toward the dayside magnetosheath plasma, which is a feature also seen, although not as clearly, during the first flyby.   

During Solar Orbiter's third flyby (Figure \ref{3venus}e,f), the ambient solar wind conditions were rather extreme with high speed and low density (about 900 km/s and 7 cm$^{-3}$, respectively, both before and after the flyby), but a normal upstream magnetic field strength \textcolor{black}{($\sim$8 nT) and normal magnetosonic Mach number $\sim$6.3, if using the ion temperature of 50 eV measured by the Proton and Alpha Analyzer/Solar Wind Analyser (PAS/SWA) on Solar Orbiter and assuming the electron temperature to be the same. The Mach number would increase to $\sim7.6$} if using an electron temperature of 10 eV instead. For comparison, during Solar Orbiter's first two flybys the Mach number was around 5.5 for an \textcolor{black}{ion} temperature of 10 eV. The trajectory during Solar Orbiter's third flyby was also somewhat different from the first two flybys, with Solar Orbiter approaching Venus' tail at a steeper angle to the X$'$-axis. At 00:22 UTC on 4 Sep 2022, there is a clear shock crossing seen in both the magnetic field and density data, with an apparent movement of the BS shortly after,  causing the spacecraft to be out in the solar wind again for about 2 minutes before crossing the shock inbound again. The time series of electron density is different from the previous flybys in that there is a local minimum in density after the entry, at around 00:33 UTC, but there are also density fluctuations following this, lasting until around 00:55 UTC, which are similar to the first two flybys. \textcolor{black}{At this time, the large magnetic field fluctuations also stop and there is an increase in the magnitude and a small rotation of the field direction.} However, this interval during the first half of the third flyby (from the inbound BS crossing until 00:55 UT) is probably due to the extreme solar wind and its variations at this time, causing large-scale variations in the magnetosheath plasma properties. While the spacecraft crossed the IMB during both the first and second flybys and entered the proper magnetotail, the third flyby did not go as deep into the tail, and the IMB was not crossed during this flyby.

\subsection{BepiColombo}
\begin{figure*}
\includegraphics[width=15cm]{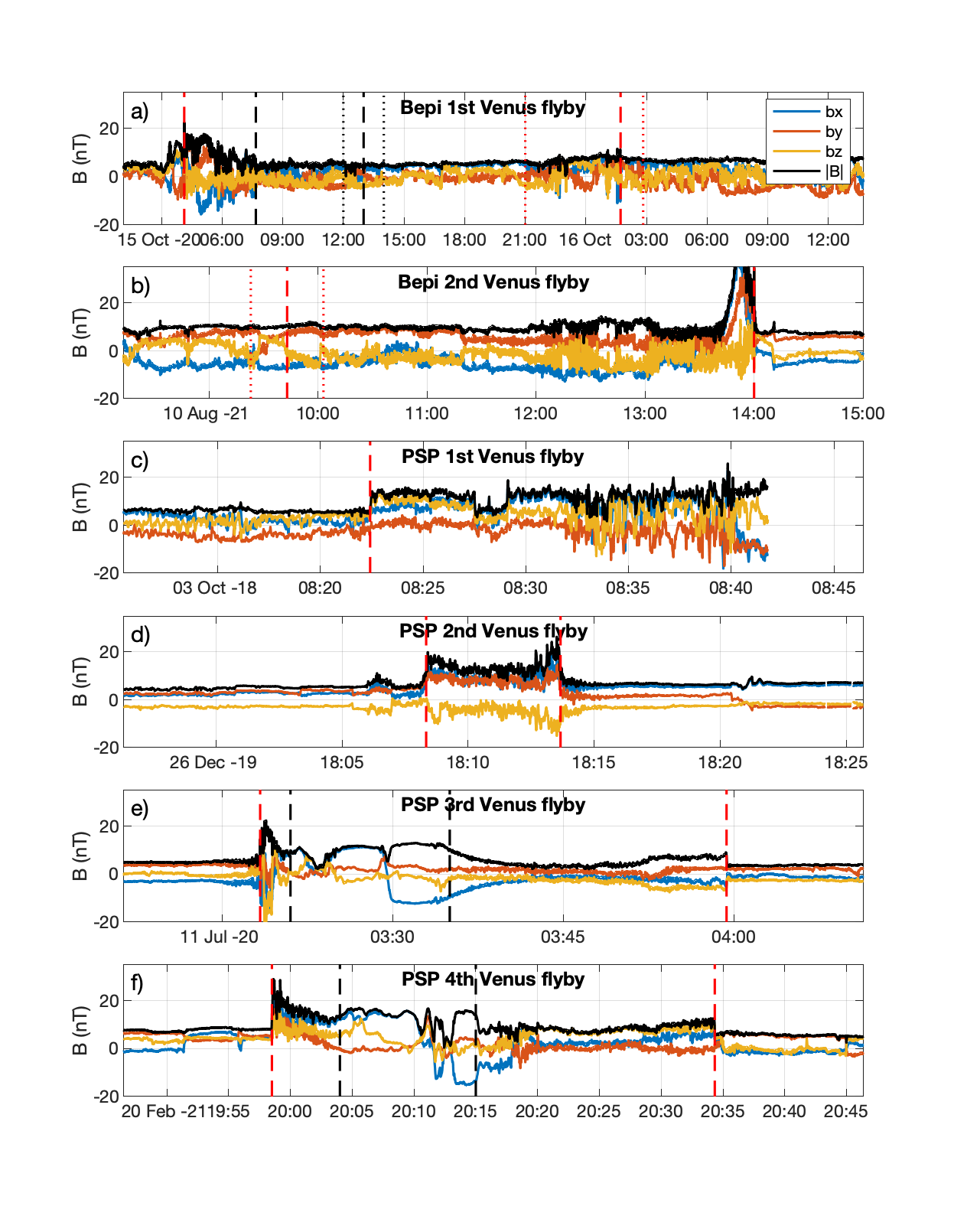}
\caption{Time series of BepiColombo and PSP magnetic field data from their Venus flybys, similar to Figure \ref{3venus}. The crossings of the BS and IMB are indicated by red and black dashed lines, respectively, \textcolor{black}{and the uncertainty intervals for some of the crossings are marked with dotted lines in panel a and b.}}
\label{2venus}
\end{figure*}

BepiColombo flew past Venus twice, on 15 Oct 2020 and 10 Aug 2021. It had trajectories similar to Solar Orbiter's first two flybys and passed through the far downtail region, but outbound during the first flyby while inbound during the second flyby. Figure \ref{2venus}a and Figure \ref{2venus}b show time series of the magnetic field measurements during these two flybys. There were no reliable plasma measurements available from BepiColombo to aid the interpretations of boundary crossings downtail. \citeA{volwerk2021b} and \citeA{aizawa2022} studied these two flybys in detail, focusing on various physical processes occurring in the plasma environment, and possible entries and exits were briefly mentioned in those papers.

During the first flyby, the far downtail IMB and BS crossings (on the outbound leg) \textcolor{black}{were challenging to identify as there was no clear or sharp transition, but rather a gradual change into the solar wind. We therefore mark the possible times/locations of these boundaries and add an uncertainty interval (see dotted lines in Fig \ref{2venus}a) for both the IMB and BS.  This interval roughly coincides with where models would suggest these two boundaries would actually be located, which is somewhat reassuring}. The entry, on the ram-side, was also somewhat unusual as the solar wind was again unsteady at this time. However, a clear shock signature is visible on the dayside at the expected location, when zooming in.

The entry through the BS during the second flyby was also challenging to identify. The start of increased fluctuations seems to be the main signature in this case, and the lack of any similar signature in the upstream solar wind measurements by Solar Orbiter \textcolor{black}{(see discussion below)} speaking in favour of a spatial change rather than a temporal \textcolor{black}{change. The inbound BS is again added with an uncertainty interval using dotted lines}. The exit showed more normal signatures of a BS crossing in the magnetic field data. BepiColombo did not go near the expected location of the tail IMB during this flyby.

\begin{figure*}
\centering
\includegraphics[width=14cm]{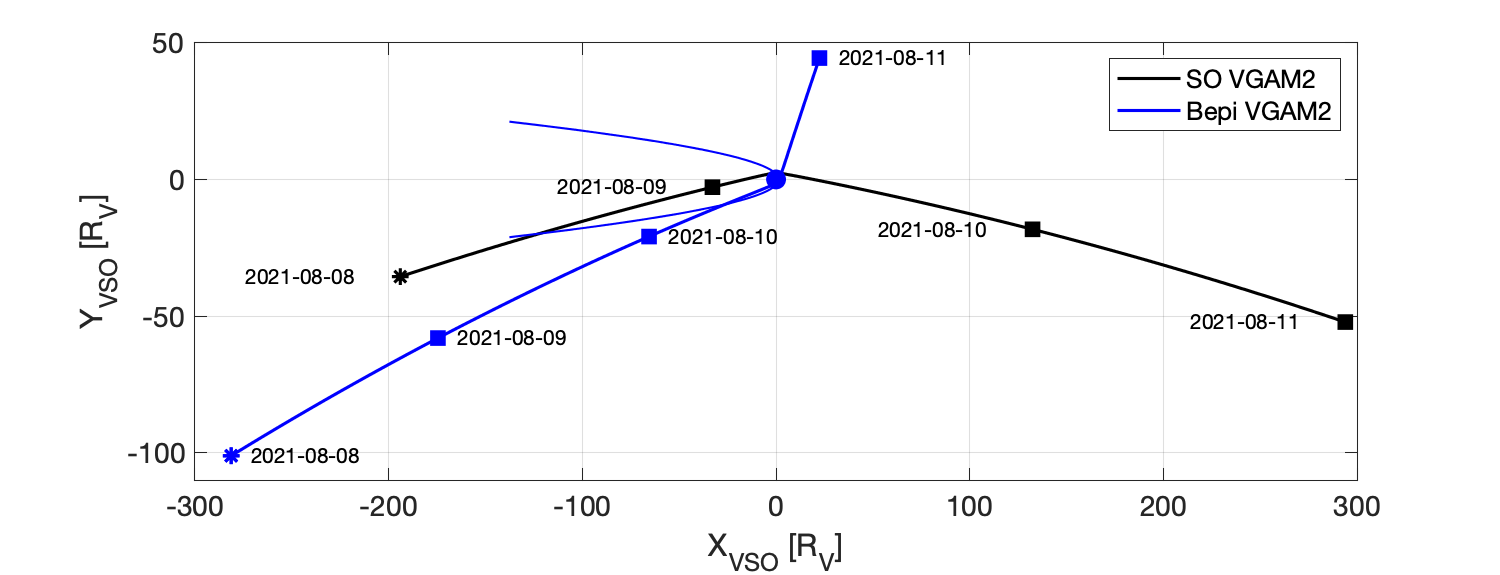}
\caption{\textcolor{black}{The trajectories of Solar Orbiter and BepiColombo during their respective 2nd flyby, which took place within a day of each other. The time covers three days, and the magnetic field data gathered during this time is shown in Figure \ref{comp}.}}
\label{traj2}
\end{figure*}

\begin{figure*}
\centering
\includegraphics[width=18cm]{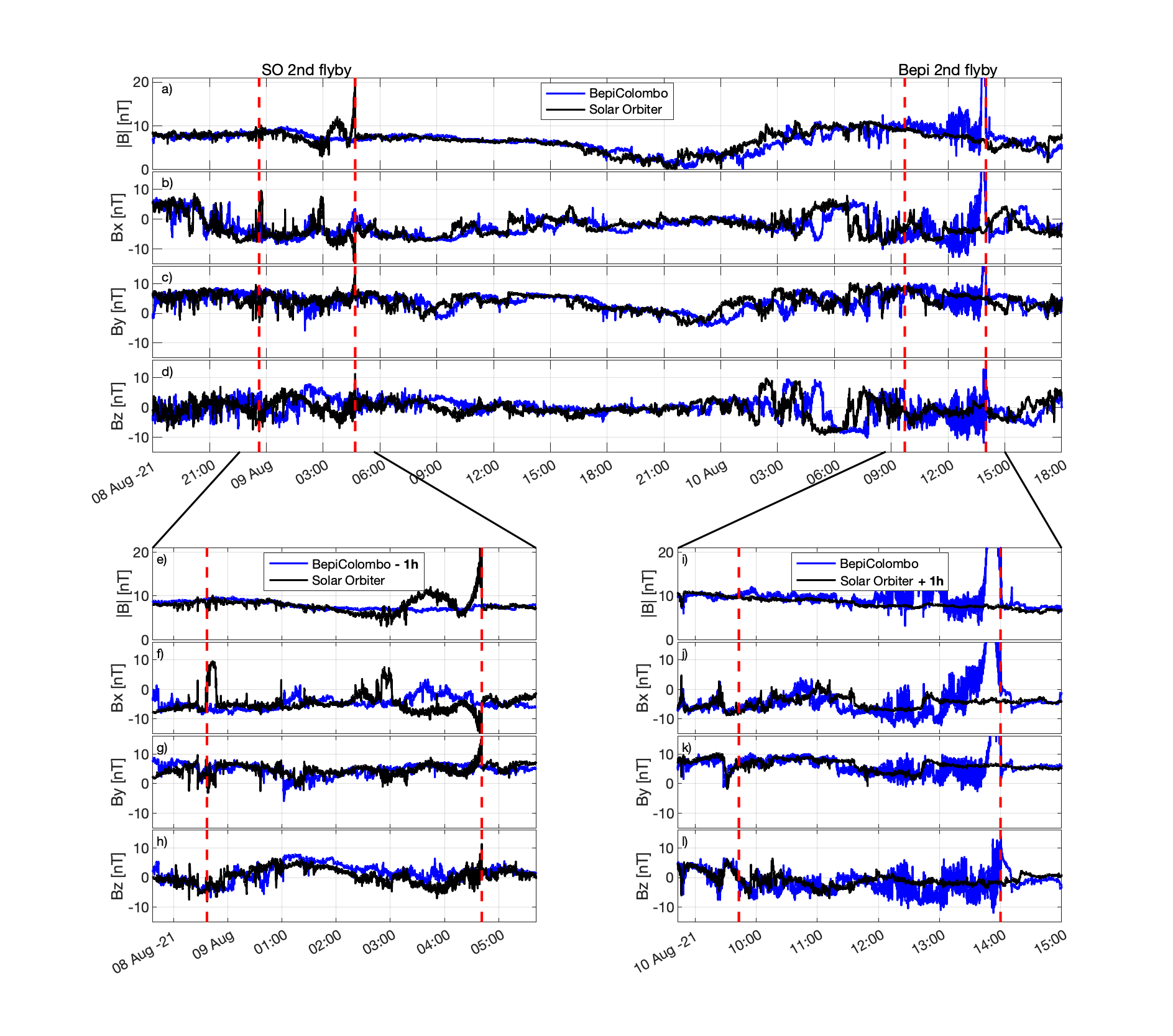}
\caption{Time series of magnetic field measurements by Solar Orbiter and BepiColombo (a-d) during their respective second Venus flybys, which took place roughly one day apart. This corresponds to a solar wind travel time from Solar Orbiter to BepiColombo of about 1 hour. The intervals when one of the spacecraft is within Venus BS are indicated by red dashed lines and a zoom in is provided in the lower panels, where in panel e-h the BepiColombo data has been shifted to -1h to the time frame of Solar Orbiter and in panels i-l the Solar Orbiter data has been shifted +1h to the time frame of BepiColombo. Note the overall apparent correlation between the two measurements, to some extent also visible within the plasma environment of Venus albeit with some differences in between the two data sets.}
\label{comp}
\end{figure*}

As mentioned above, the second Solar Orbiter and second BepiColombo flybys \textcolor{black}{fortuitously took place within a day of each other. Their trajectories during three days covering the flybys are shown in Figure \ref{traj2} in the VSO X-Y plane.}
In Figure \ref{comp}, we show a longer time series of magnetic field data from these two missions, which cover both their Venus flybys. From this, one can see how the solar wind variations clearly propagate into the Venus plasma environment, making it challenging to separate solar wind variations from plasma boundary crossings. We note that when one spacecraft is in the solar wind and the other in the far magnetotail, the ratio between the IMF and the magnetic field strength in the far tail is approximately 1. As the density is generally increased in this region (Figure \ref{3venus}), the pressure balance must be \textcolor{black}{controlled by a lower density of higher-energy plasma (not observable by RPW, and the adequate particle instruments were off at this time)}.  Overall, there is a good agreement between the two data sets when they are separated by up to 1h of solar wind travel time, and up to 100 km difference in transverse direction (from around 9 Aug 2021 to 11 Aug 2021). Solar wind structures smaller than 100 km generally do exist. Still, the signatures that are not seen in both data sets are in many cases a result of one spacecraft being in the solar wind while the other in the Venus plasma environment. For instance, Solar Orbiter's and BepiColombo's inbound BS crossings would both have been challenging to identify from each data set alone.   

\subsection{PSP}
While both Solar Orbiter and BepiColombo had their flybys going through the far tail, PSP crossed the BS within 7 R$_V$ downstream of the planet during all four flybys. At this distance, the BS is well developed and easily identifiable in magnetic field data alone (see Figure \ref{2venus}c,d,e,f). PSP stayed within the planetary plasma environment for a much shorter time than Solar Orbiter and BepiColombo, on the order of 10's of minutes. The first two flybys approached Venus from the tail and crossed the BS but not the IMB since \textcolor{black}{PSP did not} go close enough to the planet. During PSP's first flyby, the instrument was turned off in the middle of the flyby, and data was only provided from the inbound leg. The third and fourth flybys passed Venus on the nightside and crossed the BS and the IMB both inbound and outbound, as well as the tail current sheet \cite{collinson2021,collinson2022}. The BS crossings are much clearer at these relatively small distances compared to the far tail. The crossing of the IMB during PSP 3 and 4 occurred on the nightside and closer to the planet than Solar Orbiter and BepiColombo but showed a similar increase in the field strength. The exact time of the crossing was somewhat uncertain during the outbound leg of PSP 3 and during the inbound leg of PSP 4, since the transition from the sheath to the magnetotail was rather gradual in these cases.

\begin{figure*}
\includegraphics[width=14cm]{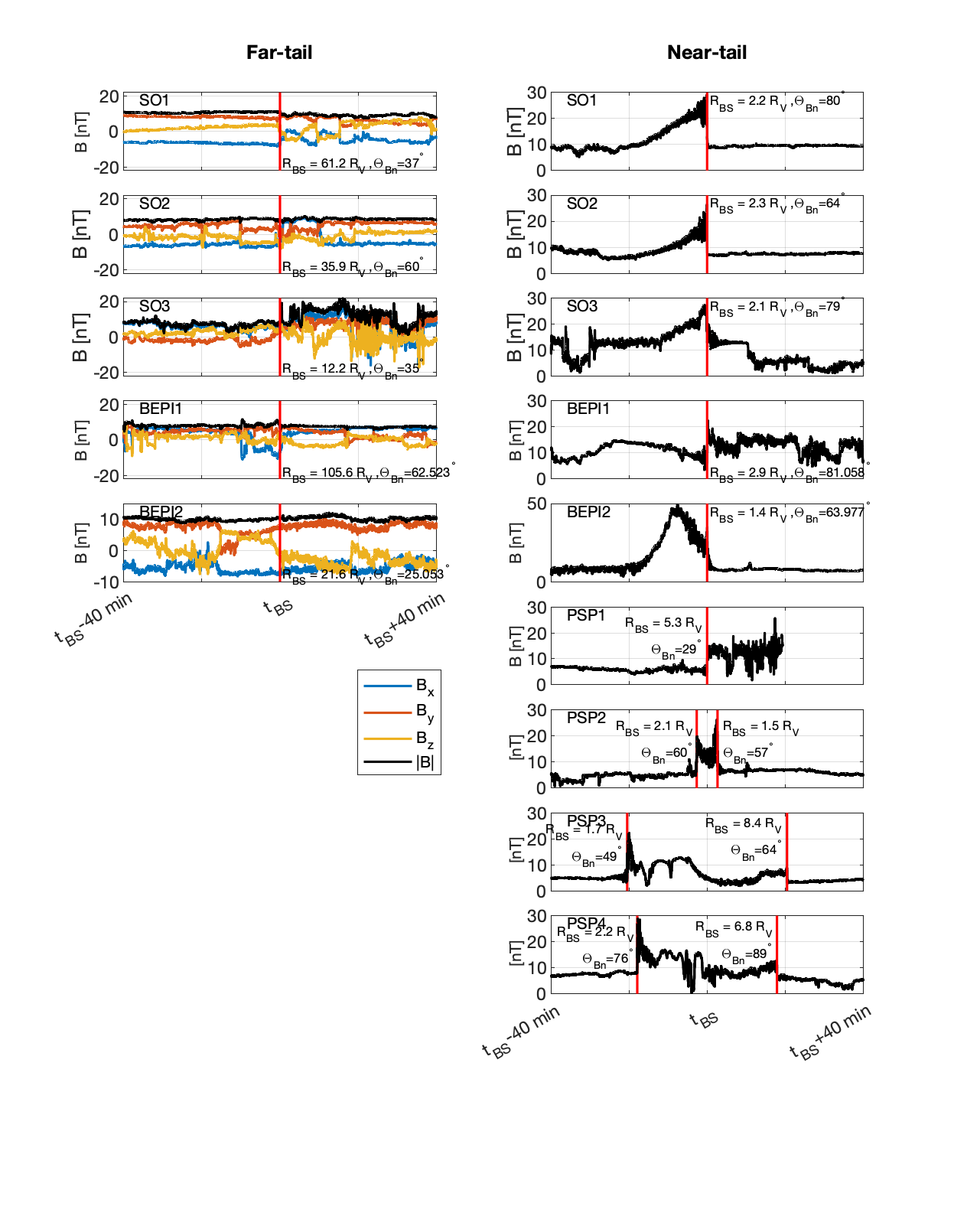}
	\caption{Time-series of magnetic field data around the time of each BS crossing, separated into the far tail and near tail region. The distance to Venus is indicated in each panel. For the panels with two PSP bow shock crossings t$_{BS}$ is set at the middle of the two. Note the large difference in structure between near tail and far tail BS.}
 \label{magdata}
\end{figure*}

In Figure \ref{magdata}, we summarize the magnetic field signatures of all BS crossings from Solar Orbiter, BepiColombo, and PSP, and separate them into the near and far tail defined as inside or outside of 10 R$_V$. We also calculate the angle \textcolor{black}{ $\theta_{Bn}$ between the local shock normal (as determined using the coplanarity theorem) and the magnetic field, which} is indicated in the figure for each BS crossing. There is a difference in the structure of a quasi-parallel or quasi-perpendicular shock, with more fore-shock waves present in the parallel case \cite{eastwood2005}. This causes some uncertainty in the actual position of the shock, but not significant for the purposes of this paper. 

\section{Boundary models}
\label{section:model}
The locations in the space around Venus corresponding to the BS and IMB crossings are shown in Figure \ref{orbit}. As mentioned above, some of the passages from the solar wind regime into the Venus plasma environment are not easily identifiable. Nevertheless, even allowing for an uncertainty of several hours, the far-down tail crossings/transitions are located well within the previous BS model of \citeA{signoles2023}, which is also included in the figure for comparison. We compare our data to this model far down-tail but it should be noted that the model was developed based on crossings in the near-tail region, i.e. within 5 R$_V$, and therefore not necessarily applicable at these distances and which warrants a new model.

\subsection{BS}
\begin{figure*}
\includegraphics[width=18cm]{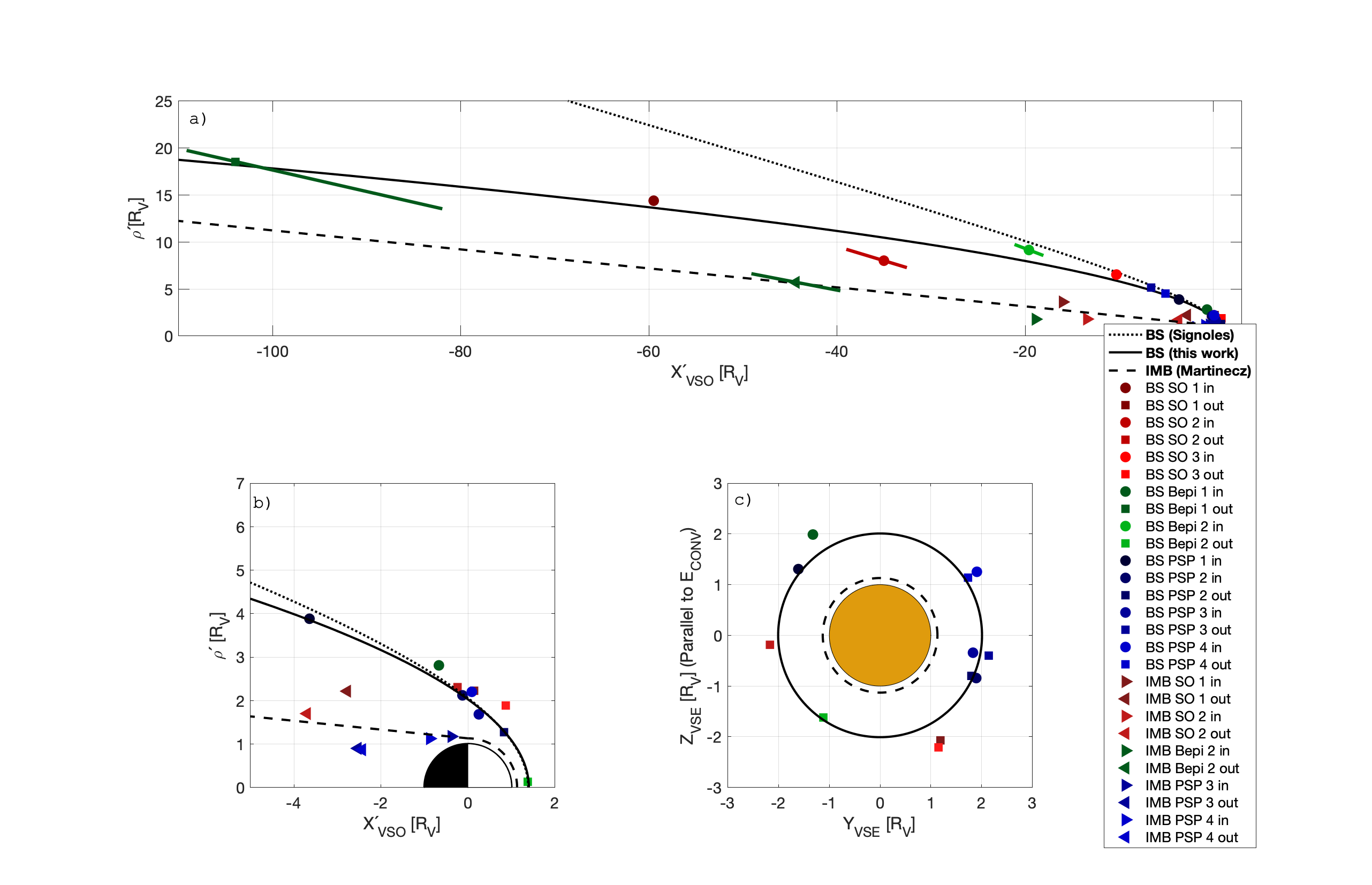}
	\caption{The locations of the crossing of the BS inbound (circles) and outbound (squares) and IMB inbound (right triangles) and IMB outbound (left triangles) by Solar Orbiter, BepiColombo and PSP shown in (a) \textcolor{black}{cylindrical aberrated VSO coordinates and (b) cylindrical aberrated VSO coordinates} zoomed in. In panel a, the approximate locations of \textcolor{black}{some of the boundary crossings are also accompanied by an uncertainty interval (solid colored line) corresponding to the intervals bound by the dotted lines in Fig \ref{3venus} and \ref{2venus}, as these exact boundary locations were challenging to accurately identify in the data.} The solid black line show the adjusted BS model from this study, the dotted line the BS model by\citeA{signoles2023} and the dashed line the IMB model from \citeA{martinecz2009}. In panel c the BS crossings from inside of 10 R$_V$ from Venus are shown in VSE coordinates in the Y$_{VSE}$-Z$_{VSE}$ plane after having been extrapolated to the terminator plane (x=0) along the model surface.}
 \label{orbit}
\end{figure*}

The statistical sample of BS crossings is poor beyond 10 R$_V$, with only four clear crossings, but as these are the only measurements we have, and which we will have even in the foreseeable future, we will use these to compile a modified model of the far-tail BS.  We use the same type of conic section model as \citeA{signoles2023} and \citeA{martinecz2009}. A least-square's fit to Equation \ref{conicsection}, with fixed values of $L=1.466$ R$_V$ and $X_0=0.688$ R$_V$, gives $\epsilon=1.001 \pm 0.002$. The error estimate corresponds  to an epsilon that increases the least-square value by 5\%. Our fit provides a smaller eccentricity with respect to the study by \citeA{signoles2023}, where $\epsilon=1.042$. \textcolor{black}{An epsilon equal to one describes a parabola while an epsilon greater than one a hyperbola. Only a hyperbola has a straight asymptote, which is expected to be the case for a Mach cone when the magnetosonic speed and flow speed are constant}. Nevertheless, such a 4\% variation in epsilon is sufficient to result in a notable displacement of the BS crossings in the tail: at 60R$_V$ the BS changes its width \textcolor{black}{(cone radius)} from 22 R$_V$ to 14 R$_V$. This fit still preserves the BS shape and location on the dayside, which we have no reason to change since the previous model was well constrained there. Using this value of the eccentricity modifies the far tail location as shown in Figure \ref{orbit}a (compare solid and dotted lines), while the near-tail region is shown in Figure \ref{orbit}b. 
\subsection{IMB}
The number of IMB crossings is equally poor with only 3 crossings beyond 10 R$_V$. A straight-line least-squares fit would not be very meaningful if not also including the crossings closer to the planet. However, the pre-existing model by \citeA{martinecz2009} actually agrees very well with such a fit with only a minor difference in the slope ($dX'/d\rho=-0.097$ instead of -0.101) and offset (1.10 R$_V$ instead of 1.13 R$_V$). This is a $\sim$3-4\% variation in the parameters but for a straight line this does not make any substantial difference: at 20 R$_V$ the position of the IMB only changes by 0.1 R$_V$.

\begin{figure*}
\includegraphics[width=15cm]{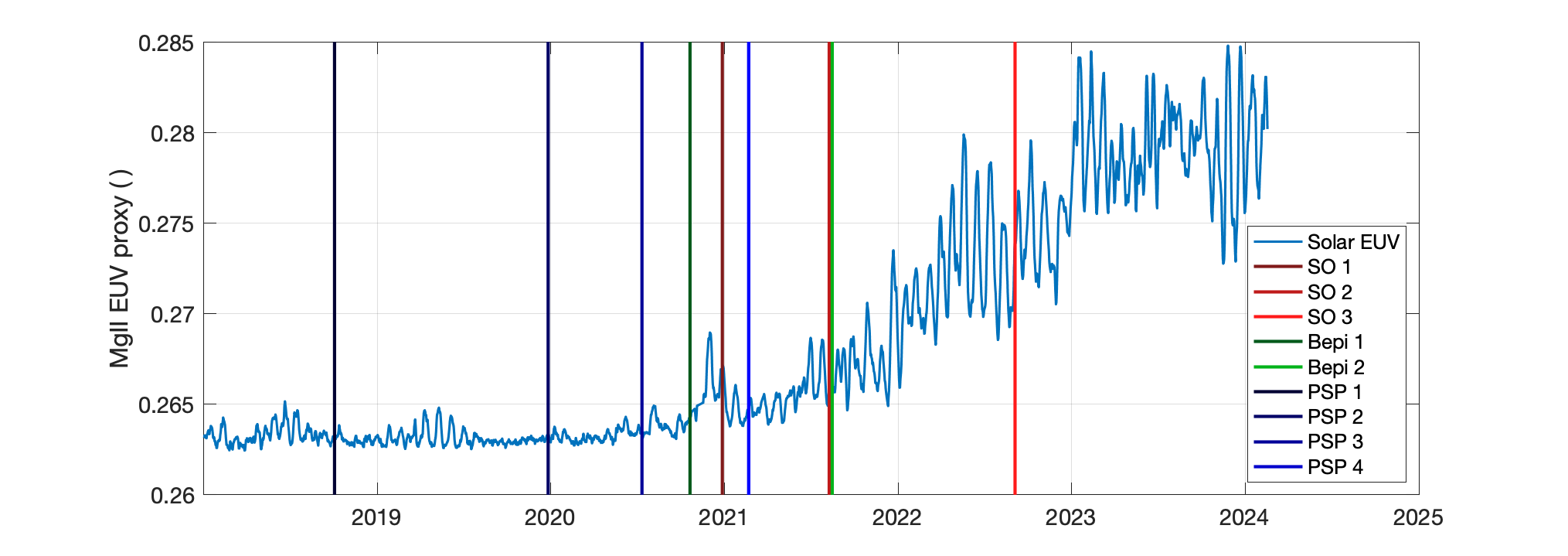}
	\caption{The times of the Venus flybys indicated on the solar cycle variations. The dimensionless solar EUV proxy is taken from observations of the MgII line at Earth orbit \cite{dudok2009}.}
 \label{euv}
\end{figure*}

\section{Discussion}\label{section:discussion}
The observed crossings of the BS and IMB by the Solar Orbiter, BepiColombo, and PSP spacecraft provide new insights into the extent of the induced magnetotail of Venus. \textcolor{black}{The farthest BS crossing, besides an uncertain crossing during BepiColombo's first flyby, was observed by Solar Orbiter at 60 R$_V$ downstream of Venus, marking a relatively clear boundary crossing. From these measurements, it can be inferred that the \textcolor{black}{BS} of Venus extends to at least 60 R$_V$ downstream and about 15-20 R$_V$ in the lateral direction this far downstream. Adjusting the previously existing empirical BS model from \citeA{signoles2023} shifts the BS about 10 R$_V$ toward the central tail at 60 R$_V$ downstream, which is a significant change.} 

Close to the planet (within 10 R$_V$), the BS is easily identifiable and so is the turbulent magnetosheath plasma within and the IMB. Farther away from Venus, the transition between solar wind plasma and Venus' plasma environment is sometimes less clear. Especially during the first BepiColombo flyby it is challenging to clearly identify the outbound IMB, and even more so during the outbound BS crossing. This ambiguity makes it debatable to claim the presence of a boundary at that time, as opposed to a gradual transition\textcolor{black}{, and we therefore include an uncertainty interval around the time of the unclear IMB and BS crossings. These intervals are chosen generously as to clearly capture the transition from one plasma regime to the next}.

\textcolor{black}{The suggested BS crossing by BepiColombo during its first flyby was at 104 R$_V$ downstream, but this crossing was rather unclear with an uncertainty interval in the range 82-109 R$_V$. The magnetic field signatures changes characteristics across this interval which suggest a transition (abrupt or not) to another plasma regime. Whether or not this is only a solar wind variation remains an open question. The inbound BS during BepiColombo's second flyby was also challenging to mark, but the upstream solar wind measurements by Solar Orbiter at this time helped identify larger changes in the magnetic field strength and a variability that could not be observed in the solar wind upstream (see Figure \ref{comp}), which speaks in favour of a boundary/spatial change.} 

\textcolor{black}{All the BS crossings are not necessarily proper shock crossings but could rather be bow waves, propagating along a Mach cone. With only a few instantaneous point measurements the actual shape of the BS is uncertain, and we really only have information on its average location. Since the upstream solar wind conditions are seldom steady on the timescales relevant here, the boundary moves and reshapes constantly. However, it is evident that the location of the BS is different from the previous empirical models. It is also worth noting that the BS model is asymptotic to a straight line when far downstream.} A least-squares fit of a straight line to the far tail crossings gives an angle of 6.5$^{\circ}$ to the X$'$-axis, intersecting the conic section model at around $X'=-10$ R$_V$. While \citeA{slavin1984} observed a Mach cone angle of 13.9$^{\circ}$ from earlier observation closer to the planet (within 15 R$_V$), \citeA{zhang2008d} and \citeA{whittaker2010} reported an angle of 10.5$^{\circ}$. Such large angles would significantly overestimate the flaring of the boundary we observe here. \textcolor{black}{We can speculate that the shape changes from the initial hyperbola when far downstream as the factors (solar wind density, temperature, velocity, magnetic field) that determine the Mach number and influencing the boundary changes with distance, and/or when the bowshock becomes more of a bow wave. Simultaneous measurements of all parameters that affect the local Mach number would be needed at the time of the crossings to more properly determine what alters the shape.}

In contrast to the BS, the location of all the IMB crossings suggests that the model of \citeA{martinecz2009} still seems to be valid in the far downtail region down to at least 20 R$_V$, without the need for adjustment of the model parameters. Although there is a considerable spread in the location of the crossings (see Figure \ref{orbit}), a least-squares fit (not shown) to these points align well with the existing model.

Previous studies \cite{russell1988, signoles2023} have shown that the BS strongly depends on the EUV flux. Figure \ref{euv} shows the timing of all flybys together with a proxy of the solar EUV flux during about half a solar cycle. All the crossings by Solar Orbiter, BepiColombo, and PSP occurred around the minimum of the solar cycle, except for the third Solar Orbiter flyby, which occurred about halfway between solar minimum and solar maximum, and was located farther out from the model surface than all of the other crossings. The third Solar Orbiter inbound crossing down the tail was also somewhat farther out than average. The role of EUV emission seems therefore consistent with the location of the BS crossings, which slowly moves outward during the increasing phase of the solar cycle. The third Solar Orbiter flyby took place during extreme solar wind conditions (speed of about 900 km/s and density around 7 cm$^{-3}$). The boundary was not compressed by this high dynamic pressure but rather found farther out, presumably due to the counter-acting increase in EUV flux. Also, if lowering the Mach number, the BS would also move outward \cite{edberg2010b}. However, the magnetosonic Mach number was around \textcolor{black}{6.3} here which is a rather typical value at Venus. Hence, the increased BS distance during this flyby is therefore possibly explained by the effect of increased solar EUV flux during the increasing phase of the solar cycle.

The trajectories of especially Solar Orbiter and BepiColombo were rather tangential to the BS and IMB surfaces in the tail. As crossings are naturally only found along the orbit it could be that it is only when the boundary is compressed to the position of the spacecraft that we observe it. The resulting boundary shape would then inherently be similar to the spacecraft path. \textcolor{black}{With the spacecraft flying past relatively fast (as opposed to having a probe dwelling at the boundary location for a long time), we cannot asses the boundary motion, or the absence thereof, with certainty. Similarly, without a proper upstream monitor we cannot asses if the upstream conditions cause the boundary to be temporarily compressed at these times. Nevertheless, we do not normally see multiple boundary crossings in the measurements presented here, which would happen if the boundary moves in and out past the spacecraft. An exception is the inbound third Solar Orbiter flyby, but then only for a brief time corresponding to a small spatial difference. This rather indicates that the boundary location, and its shape, is quite steady in space for each flyby considered here, at least on the timescale of a flyby duration}.

Other factors that could potentially influence the BS include the convective electric field. This could be manifested as a shift of the BS in the +Z$'_{VSE}$ direction, but such an effect does not appear to be significant. In Figure \ref{orbit}c, the distance of the crossings are shown in the VSE coordinate system, to illustrate if there is any significant effect of acceleration of ions by the solar wind convective electric field that would lead to a shift of the BS. In the VSE coordinate system the Y$'_{VSE}$ axis is aligned with the upstream magnetic direction projected on the Y-Z plane such that Z$'_{VSE}$ is parallel to the convective electric field ($\bf{E}=-\bf{v} \times \bf{B}$) direction. A shift of the boundary location in the vertical direction could potentially have been observed if there was a strong dependence on the convection electric field, but this does not appear to be the case, \textcolor{black}{at least not within this relatively small statistical sample.}

\section{Conclusions}
\label{section:conclusion} 

In this study, we have synthesized measurements from nine Venus flybys conducted by the Solar Orbiter, BepiColombo, and PSP missions to investigate the boundary structure of the far magnetotail of Venus. 
We refine the previously existing empirical conic section BS model to better align with the far tail crossings, with model parameters $L=1.466$ R$_V$ and $X_0=0.688$ R$_V$ unchanged while we adjusted the parameter $\epsilon=1.001 \pm 0.002$. \textcolor{black}{This narrows the Venus plasma environment and puts the BS $\sim$10 R$_V$ closer to the central tail at 60 R$_V$ downstream.} While a conic section was used for the improved model shape, a straight line representing a Mach cone at an angle of 6.5$^{\circ}$ would also adequately fit the data due to the large length scales involved. \textcolor{black}{This straight line asymptote is consistent with the main driving factor for shaping the boundary far downtail is the solar wind Mach number. Since we are only using instantaneous single-point measurements we cannot say anything definitive about the shape but rather only determine the average location. }

\textcolor{black}{The location of the BS far downtail is not found to be significantly compressed by the extreme solar wind dynamic pressure conditions during Solar Orbiter's third flyby, but rather more influence by the increasing EUV flux toward the solar maximum, which acts to expand the boundary.}

 \textcolor{black}{The pre-existing IMB model appears to remain valid up to 20 R$_V$ downstream, despite a notable spread in the individual crossings. It is presumably varying with upstream solar wind parameters as previous larger statistical studies have suggested.}

The BS crossings are always evident \textcolor{black}{in the data close to the planet. Conversely, far downtail the transition from solar wind to Venus' plasma environment is generally less clear, posing challenges in precisely pinpointing the boundary location or transition between regions. Nevertheless, signatures of Venus' plasma environment in terms of plasma boundary signatures} extend to at least 60 R$_V$ downstream.

While several flybys remain for Solar Orbiter and Parker Solar Probe, none of them will pass through the far tail again. Hence, the set of flybys included in this study provided the bulk of observations that we will have of the far magnetotail of Venus for decades to come. 

\section{Data Availability Statement}
\textcolor{black}{The Solar Orbiter data are available through ESA's Solar Orbiter archive \url{(https://soar.esac.esa.int/soar/)}. For density measurements, we used the RPW LFR data sets \cite{maksimovic2020b}. }
\textcolor{black}{BepiColombo MPO-MAG will be available through ESA’s Planetary Science Archive, but is now stored on Zenodo (\url{https://doi.org/10.5281/zenodo.13373280}) \cite{edberg2024data}. }
\textcolor{black}{Parker Solar Probe data are available on the Coordinated Data Analysis Web (CDAWeb) at NASA Goddard Space Flight Center (\url{https://cdaweb.gsfc.nasa.gov/}) \cite{pspdata}.}

\acknowledgments
 The authors acknowledge all members of the Solar Orbiter, BepiColombo and Parker Solar Probe missions for their unstinted efforts in making these missions successful. The authors extend their gratitude to SNSA (formerly SNSB) for their support of the Solar Orbiter/BIAS instrument. NE and KK gratefully acknowledge funding from the Swedish Research Council (Vetenskapsrådet) under contract 2020-03962. LSV was supported by the Swedish Research Council (VR) Research Grant N. 2022-03352. LH acknowledges the support of Center National d’Etudes Spatiales (CNES, France) to the BepiColombo and Solar Orbiter mission. DP received support from the Czech Science Foundation (grant no. 22-10775S). K.S and M.L. acknowledge support through UK-STFC grant ST/W00089X/1.  B.S.-C. acknowledge support through UK-STFC Ernest Rutherford Fellowship ST/V004115/1. D.H., U.A. and D.C. were supported by the German Ministerium f\"{u}r Wirtschaft und Klimaschutz and the German Zentrum für Luft‐ und Raumfahrt under contract 50QW2202. This research was supported by the International Space Science Institute (ISSI) in Bern, through ISSI International Team project \#23-593. 

\bibliography{refs3}

\end{document}